\documentclass[journal,article,submit,moreauthors,pdflatex,12pt,a4paper]{mdpi} 
\lastpage{x}
\doinum{10.3390/------}
\pubvolume{xx}
\pubyear{2012}
\history{Received: xx / Accepted: xx / Published: xx}

\usepackage{amsmath}
\usepackage{amssymb}

\Title{Viscosity in Modified Gravity}

\Author{Iver Brevik $^{1,\star}$ }

\address{%
$^{1}$ Department of Energy and Process Engineering, Norwegian University of Science and Technology, N-7491 Trondheim, Norway\\
}

\corres{E-Mail iver.h.brevik@ntnu.no, Tel. +47 7359 3555,  Fax +47 7359 3491.}

\abstract{A bulk viscosity is introduced in the formalism of modified gravity. It is shown that, on the basis of a natural scaling law for the viscosity, a simple solution can be found for quantities such as the Hubble parameter and the energy density. These solutions may incorporate a viscosity-induced Big Rip singularity. By introducing a phase transition in the cosmic fluid, the future singularity can nevertheless in principle be avoided.}


\begin{document}

\section{Introduction}

Modified gravity has become an active branch of modern cosmology, attempting to give a unified description of the early (inflationary) epoch of the universe and at the same time intending to account for the accelerated expansion at the later stages. Useful reviews on modified gravity theories  can be found in Refs.~\cite{nojiri07,nojiri11,bamba12}.

Most treatises on modified gravity, as well as on standard gravity, assume the cosmic fluid to be ideal, i.e. nonviscous. From a hydrodynamicist's point of view this is somewhat surprising, since there are several situations in fluid mechanics, even in homogeneous space without boundaries, where the two viscosity coefficients, the shear coefficient $\eta$ and the bulk  coefficient, $\zeta$ come into play. This means a deviation from thermal equilibrium to the first order. Such a theory means in effect acceptance of  the Eckart 1940 theory \cite{eckart40}. An important  property of the Eckart assumption is that the theory becomes noncausal. By taking into account second order deviations from thermal equilibrium, one can obtain a causal theory respecting special relativity. Pioneering articles on causal fluid mechanics are those of   M{\"u}ller \cite{muller67}, Israel \cite{israel76}, and Israel and Stewart \cite{israel79}.    A  recent review can be found in Ref.~\cite{brevik12}.  Because of the assumption about spatial isotropy, the shear coefficient is usually omitted.

Our purpose in the following will be to include the bulk viscosity $\zeta$ in the modified gravity formalism. We consider  the case when $\zeta$ is satisfying a scaling law, reducing in the Einstein case to a form proportional to the Hubble parameter. It turns out that this scaling law is quite useful. We survey first earlier developments along this line, extracting material largely from our earlier papers \cite{brevik05,brevik06,brevik08}. Thereafter, as a novel development we investigate how the occurrence of a phase transition
 can change the development of the universe, especially in the later stages approaching the future singularity. (It may here appear natural to relate such a phase transition with the onset of a turbulent state of motion.) It is shown that such a transition may in principle be enough to prevent the singularity to occur at all. This part of the paper, covered in Sect. 4, is  a generalization of the viscous/turbulent theory for standard cosmology recently given in Refs.~\cite{brevik11} and \cite{brevik12B}.


\section{Fundamental Formalism}

 The action in modified gravity is conventionally written in the general form
\begin{equation}
 S=\int d^4x \sqrt{-g}\left[\frac{F(R)}{2\kappa^2} +\cal{L}_{\rm matter}\right], \label{1}
 \end{equation}
 where $\kappa^2=8\pi G$, and where    $ \cal{L}_{\rm matter} $  is the matter Lagrangian. The equations of motion are
 \begin{equation}
 -\frac{1}{2}g_{\mu\nu}F(R)+R_{\mu\nu}F'(R)-\nabla_{\mu}\nabla_{\nu}F'(R)+g_{\mu\nu}\Box F'(R)=\kappa^2T_{\mu\nu}^{\rm matter}, \label{2}
 \end{equation}
 where $T_{\mu\nu}^{\rm matter}$ is the energy-momentum tensor corresponding to $\cal{L}_{\rm matter}$.

 We shall however in the following not consider the general case, but limit ourselves to the special form where
 \begin{equation}
 F(R)=f_0 R^\alpha, \label{3}
 \end{equation}
  $f_0$ and $\alpha$ being constants. This model has been used before, by Abdalla et al. \cite{abdalla05} and others. The case of Einstein gravity corresponds to $f_0=1$ and $\alpha=1$. This choice appears to  be natural from a mathematical viewpoint,   and in our case it will  play an important role in connection with the scaling law for the bulk viscosity; cf. Eq.~(\ref{21}) below. Quadratic Lagrangians were considered also earlier, by Barrow \cite{barrow87,barrow86}.

We  assume the spatially flat FRW metric
 \begin{equation}
 ds^2=-dt^2+a^2(t)d{\bf x}^2, \label{3A}
 \end{equation}
 and put the cosmological constant $\Lambda=0$. In comoving coordinates, the components of the four-velocity $U^\mu$ are $U^0=1, U^i=0$. Introducing the projection tensor $h_{\mu\nu}=g_{\mu\nu}+U_\mu U_\nu$ we have for the energy-momentum tensor
 \begin{equation}
 T_{\mu\nu}=\rho U_\mu U_\nu+\tilde{p} h_{\mu\nu}, \label{4}
 \end{equation}
 where $\tilde{p}$ is the effective pressure
 \begin{equation}
 \tilde{p}=p-3H\zeta. \label{4A}
 \end{equation}
 The scalar expansion is $\theta= 3\dot{a}/a=3H$, with $H$ the Hubble parameter. The shear viscosity is here omitted.

 The equations of motion following from the above action are
 \begin{equation}
  -\frac{1}{2}f_0g_{\mu\nu}R^\alpha +\alpha f_0R_{\mu\nu}R^{\alpha-1}-\alpha f_0\nabla_\mu\nabla_\nu R^{\alpha-1}
 +\alpha f_0g_{\mu\nu}\Box R^{\alpha-1}=\kappa^2T_{\mu \nu}^{\rm matter}. \label{5}
 \end{equation}

 The equation of state for the fluid is written as
 \begin{equation}
 p=w\rho \equiv (\gamma-1)\rho. \label{6}
 \end{equation}
 If $w=-1$ or $p=-\rho$ the fluid is a vacuum fluid with strange thermodynamical properties such as negative entropies (cf., for instance, Ref.~\cite{brevik04}). Recent observations indicate that $w=-1.04^{+0.09}_{-0.10}$ \cite{nakamura10,amanullah10}. It has been conjectured that $w$ is a function varying with time, perhaps even oscillatory, and that  $w$ might have been around 0 at redshift $z$  of order unity  \cite{vikman07}. The quintessence region $-1<w<-1/3$, and the phantom region region $w<-1$, are both of physical interest. Both quintessence and phantom fluids imply the inequality $\rho+3p \leq 0$, thus breaking the strong energy condition.

 We now consider the (00) component of Eq.~(\ref{5}), observing that $R_{00}=-3\ddot{a}/a$ and $R=6(\dot{H}+2H^2)$. With $T_{00}^{\rm matter}=\rho$ we obtain
 \begin{equation}
 \frac{1}{2}f_0R^\alpha -3\alpha f_0(\dot{H}+H^2)R^{\alpha-1}+3\alpha (\alpha-1)f_0HR^{\alpha-2}\dot{R}=\kappa^2\rho. \label{7}
 \end{equation}
 An important property of (\ref{7}) is that the four-divergence of the LHS is equal to zero, $\nabla^\nu T_{\mu\nu}^{\rm matter}=0$ \cite{koivisto05}. This is as in Einstein's gravity, meaning that conservation of energy-momentum follows from the field equations. The energy conservation equation becomes
 \begin{equation}
 \dot{\rho}+(\rho+p)3H=9\zeta H^2. \label{8}
 \end{equation}
 Differentiating (\ref{7}) with respect to $t$  and  inserting $\dot \rho$ from (\ref{8}), we get
 \[ \frac{3}{2}\gamma f_0R^\alpha +3\alpha f_0[2\dot{H}-3\gamma (\dot{H}+H^2)]R^{\alpha-1}+3\alpha (\alpha-1)f_0[(3\gamma-1)H\dot{R}+\ddot{R}]R^{\alpha-2} \]
 \begin{equation}
 +3\alpha (\alpha-1)(\alpha-2)f_0\dot{R}^2R^{\alpha-3}=9\kappa^2\zeta H. \label{9}
 \end{equation}
 Inserting $R=6(\dot{H}+2H^2)$ we see that this equation for $H(t)$ is quite complicated. We shall be interested in solutions related to the future singularity, and make therefore the ansatz
 \begin{equation}
 H=\frac{H_0}{X}, \quad {\rm{where}} \quad X=1-BH_0t, \label{10}
 \end{equation}
 where $H_0$ is the Hubble parameter at present time, and $B$ a non dimensional constant. If a future singularity is to happen, $B$ must be positive.

 Before closing this section, it is desirable to comment on stability issues for our ansatz (\ref{3}) for the modified Lagrangian. A theory of modified gravity should admit an asymptotically flat, static spherically symmetric solution. Now, we expect that  the expression (\ref{2}) for the complete action, with (\ref{3}) inserted, will not be the full solution. It is reasonable to expect that the modified part will contain also other terms so that (\ref{2}) makes up only a part of the complete action. Nevertheless, it is of interest to ask to what extent (\ref{3}), when take separately, will behave with respect to the stability requirements. In the Solar system, far from the sources, it is known that $R \approx 10^{-61}$ eV$^2$; it corresponds to one one hydrogen atom per cubic centimeter.
 [Note that 1 eV= $5.068 \times 10^4$ cm$^{-1}$.] On a planet, $R=R_b \approx 10^{-38}$ eV$^2$, whereas the average curvature in the universe is $R \approx 10^{-66}$ eV$^2$. According to the stability analysis of Elizalde {\it et al.} \cite{elizalde11}, the stability condition for matter is
 \begin{equation}
 F''(R_b) >0, \quad {\rm where} ~~R_b \approx 10^{-38}~{\rm eV}^2. \label{10A}
 \end{equation}
 In our case, this means merely that the exponent $\alpha$ in the expression (\ref{3}) has to be greater than one. The stability condition on  $\alpha$ is quite modest.


\section{Special Cases}

It is now mathematically simplifying, and physically instructive, to focus on special cases.

\subsection{Einstein Action}

 As mentioned above, Einstein's gravity corresponds to   $f_0=1$ and $\alpha =1$. It means that the Lagrangian is linear in $R$. It is natural to consider this case as a reference case before embarking on the nonlinear general situation.

 We first have to adopt a definite form for the bulk viscosity. The simplest choice would be to put $\zeta =$ constant. There are however reasons to assume a slightly more complicated form, namely to put $\zeta$ proportional to the Hubble parameter $H$. This is physically natural, in view of the large fluid velocities expected near the future singularity. Such violent conditions should correspond to an increased value of $\zeta$. We shall take $\zeta$ to be proportional to the scalar expansion, $\theta=3H$,
 \begin{equation}
 \zeta =3\tau_E H, \label{11}
 \end{equation}
 $\tau_E$ being the proportionality constant in the Einstein theory. An important property of this particular form, shown in Ref.~\cite{brevik05B}, is that if $\tau_E$ is sufficiently large to satisfy the condition
 \begin{equation}
 \chi \equiv -\gamma+3\kappa^2\tau_E >0, \label{12}
 \end{equation}
 then a Big Rip singularity is encountered after a finite time $t$. Even if the universe starts out from the quintessence region $(-1<w<-1/3)$ or $(0<\gamma<2/3)$, the presence of a sufficiently large bulk viscosity will drive it into the phantom region $(w<-1)$ and thereafter inevitably into the Big Rip singularity.

 From the governing equations we get
 \begin{equation}
 B=\frac{3}{2}\chi, \quad H_0=\sqrt{\frac{1}{3}\kappa^2\rho_0}, \label{13}
 \end{equation}
 where $\rho_0$ is the present ($t=0$) value of the energy density. The time dependent value $\rho_E$ for the energy density according to the Einstein theory becomes
 \begin{equation}
 \rho_E=\frac{\rho_0}{X^2}. \label{14}
 \end{equation}
 We ought here to mention that other forms for the bulk viscosity,  more complicated than the form (\ref{11}) above, have been suggested. One possibility is that $\zeta$, in addition to the term proportional to $H$, contains also a term proportional to $\ddot{a}/a$. See further discussions on this topic in Refs.~\cite{ren06} and \cite{mostafapoor11}.

 \subsection{Modified Gravity Action}

 Consider now the modified gravity fluid, for which $f_0$ and $\alpha$ are arbitrary constants. As before, we look for solutions satisfying the ansatz (\ref{10}). It turns out that such solutions exist, if we model the bulk viscosity $\zeta_\alpha$ according to the following scaling law \cite{brevik06,brevik06B},
 \begin{equation}
 \zeta_\alpha =\tau_\alpha \theta^{2\alpha-1} =\tau_\alpha (3H)^{2\alpha-1}. \label{15}
 \end{equation}
 We see that this scaling fits nicely with our results from the preceding subsection: if $\alpha=1$, our previous form (\ref{11}) follows. The time-dependent factors in (\ref{9}) drop out, and we get the following equation determining $B$,
 \[
 (B+2)^{\alpha-1}\left\{ 9(2-\alpha)\gamma +3[\alpha+3\gamma+\alpha (2\alpha-3)(3\gamma-1)]B +6\alpha(\alpha-1)(2\alpha-1)B^2\right\} \]
 \begin{equation}
 = \frac{18\kappa^2}{f_0}\left(\frac{3}{2}\right)^\alpha \tau_\alpha. \label{16}
 \end{equation}
 This equation is in general complicated. Let us consider $\alpha=2, \gamma=0$ as a typical example (recall that  $\gamma=0$ corresponds to a vacuum fluid). Then we obtain from (\ref{16}) $( \tau_\alpha \rightarrow \tau_2)$,
 \begin{equation}
 B^3+2B^2-\frac{9\kappa^2\tau_2}{8f_0} = 0. \label{17}
 \end{equation}
 If the LHS is drawn as a function of $B$, it is seen that there is a local maximum at $B=-4/3$ and a local negative minimum at $B=0$, irrespective of the value of $\tau_2$. For all positive $\tau_2$ there is thus one single positive root. This root is viscosity-induced, and leads to the Big Rip singularity. When $\tau_2$ increases from zero, there is a parameter region in which there are three real roots. Assume this region, and introduce an angle $\phi \in [0, 180^0]$ such that
 \begin{equation}
 \cos \phi=-\left( 1-\frac{243}{128}\frac{\kappa^2\tau_2}{f_0}\right). \label{18}
 \end{equation}
 Then the actual value of the root can be expressed as
 \begin{equation}B=-\frac{2}{3}+\frac{4}{3}\cos \left( \frac{\phi}{3}+240^0 \right). \label{19}
 \end{equation}
 For instance, if we choose $\phi=120^0$, the positive solution becomes $B=0.3547$. According to Eq.~(\ref{10}) this gives the following Big Rip time
 \begin{equation}
 t_{BR}=\frac{1}{B}\frac{1}{H_0}=\frac{2.819}{H_0}. \label{19A}
 \end{equation}

 We may also note the  general relation for $B$ following from the energy conservation equation (\ref{8}), when $\rho \rightarrow \rho_\alpha, p\rightarrow p_\alpha, \zeta \rightarrow \zeta_\alpha$,
 \begin{equation}
 B=-\frac{3\gamma}{2\alpha}+\frac{3\tau_\alpha}{2\alpha}\frac{(3H_0)^{2\alpha}}{\rho_0}. \label{20}
 \end{equation}
 Here we used
 \begin{equation}
 \zeta_\alpha=\tau_\alpha \left(\frac{3H_0}{X}\right)^{2\alpha-1}, \quad \rho_\alpha=\frac{\rho_0}{X^{2\alpha}}, \label{21}
 \end{equation}
 and for simplicity we used the same initial conditions at $t=0$ for the modified fluid as for the Einstein fluid, $\rho_{0\alpha}=\rho_{0E} \equiv \rho_0$, and $H_{0\alpha}=H_{0E} \equiv H_0$.


\section{On the Possibility of a Phase Transition in the Late Universe}

In the preceding we have  surveyed bulk viscosity-induced generalizations of modified gravity, following essentially the earlier treatments in Refs.~\cite{brevik06,brevik06B,brevik08}. Our intention in the following, as a new contribution, will be to discuss the flexibility that the above model possesses with respect to sudden changes in the time development (we will refer to it as phase transitions) in the late universe. The main point is the different solutions for $B$ in the governing equation (\ref{9}) that are possible when the scaling  ansatz (\ref{15}) is inserted. We obtain the following algebraic equation for $B$,  for definiteness still assuming $\alpha=2$,
\begin{equation}
B^3+(2+\frac{3}{4}\gamma)B^2+\frac{3}{2}\gamma B-\frac{9}{8}\frac{\kappa^2 \tau_2}{f_0}=0. \label{22}
\end{equation}
This equation generalizes (\ref{17}) to the case of nonvanishing $\gamma$.

Consider the following scenario: the universe starts out from present time $t=0$ and follows the equations of modified gravity, with a $\tau_2$-induced bulk viscosity corresponding to a positive value of $B$. That means, the universe develops according to
\begin{equation}
H=\frac{H_0}{X}, \quad \zeta_2=\tau_2\left(\frac{3H_0}{X}\right)^3, \quad \rho_2=\frac{\rho_0}{X^4}, \label{23}
\end{equation}
with $X=1-BH_0t$.  The universe thus enfaces a future singularity at large times. Let now, at a fixed time that we shall call $t_*$, there be a phase transition in the cosmic fluid implying that the effect from $\tau_2$ goes away. It means that the further development of the fluid will be determined by the  $\gamma$-dependent roots of Eq.~(\ref{22}) when $\tau_2=0$. There are three roots:

\noindent 1) The first is $B=0$. This is the de Sitter case, corresponding to
\begin{equation}
H=H_*, \quad \rho_2=\rho_*, \label{24}
\end{equation}
where  $H_*$ and $\rho_*$  follow from (\ref{23}) when $t=t_*$.
By assuming that $|\gamma| \ll 1$, which is of main physical interest, we see that it is easy to determine the remaining two roots. One of them is

\noindent 2) $B=-2$. This means
\begin{equation}
H=\frac{H_*}{1+2H_*(t-t_*)}, \quad \rho=\frac{\rho_*}{[1+2H_*(t-t_*)]^2}. \label{25}
\end{equation}
The accelerated expansion is accordingly reversed at $t=t_*$, and the density goes smoothly to zero at large times.

\noindent 3) The third root is $B=-3\gamma/4$, which yields
\begin{equation}
H=\frac{H_*}{1+\frac{3}{4}\gamma H_*(t-t_*)}, \quad \rho=\frac{\rho_*}{[1+\frac{3}{4}\gamma H_*(t-t_*)]^2}. \label{26}
\end{equation}
The sign of $\gamma$ is important here. If the equation-of-state parameter $w$ lies in the quintessence region, $w>-1$ ($\gamma>0$), then the density of the universe will go to zero for large times, like for the case 2) above. By contrast, in the phantom region $w<-1$ ($\gamma<0$), the universe will actually move towards a Big Rip again, although very weakly so.

Finally, it is of interest to compare the above results with those obtained in ordinary viscous cosmology when the universe, similarly as above,  is thought to undergo a phase transition at a definite time $t_*$. Such an investigation was recently carried out in Ref.~\cite{brevik12B} (the one-component  case treated in Sect. VI). Consider the following model: the universe starts from $t=0$ as an ordinary viscous fluid with a constant bulk viscosity,
\begin{equation}
\zeta={\rm constant}\equiv \zeta_0, \label{27}
\end{equation}
and develops according to the Friedmann equations. Assume that the universe is in the phantom region, $\gamma<0$. It follows that in the initial period $0<t<t_*$,
\begin{equation}
H=\frac{H_0 e^{t/t_c}}{1-\frac{3}{2}|\gamma|H_0t_c(e^{t/t_c}-1)}, \label{28}
\end{equation}
\begin{equation}
\rho=\frac{\rho_0 e^{2t/t_c}}{ [1-\frac{3}{2}|\gamma|H_0t_c(e^{t/t_c}-1)]^2}, \label{29}
\end{equation}
where $t_c$ means the "viscosity time",
\begin{equation}
t_c=\left(\frac{3}{2}\kappa^2\zeta_0\right)^{-1}. \label{30}
\end{equation}
According to these equations the universe develops towards a Big Rip.
Now, after  $t=t_*$ we imagine an era for which
 $\gamma_{\rm turb}=1+w_{\rm turb} >0$ and an equation of state of the form
\begin{equation}
p_{\rm turb}=w_{\rm turb}\rho_{\rm turb}.  \label{31}
\end{equation}
Here the subscript "turb" refers to our association in \cite{brevik12B} of the transition at $t=t_*$ into an era dominated by turbulence.

Then, for $t>t_*$,
\begin{equation}
H=\frac{H_*}{1+\frac{3}{2}\gamma_{\rm turb}H_*(t-t_*)}, \label{32}
\end{equation}
\begin{equation}
\rho=\frac{\rho_*}{[1+\frac{3}{2}\gamma_{\rm turb}H_*(t-t_*)]^2}. \label{33}
\end{equation}
This means a dilution of the density again, at large times. The Big Rip may thus be avoided, as a result of a phase transition in the cosmic fluid. We see that in this sense the behavior is similar in the two cases, modified or ordinary, gravity.

It ought to be made clear that we do not at present have a specific model of the phase transition suggested at $t=t_*$.  Our association with a turbulent state of motion is however quite natural, on the basis of the following consideration: In states of violent local motions of the cosmic fluid near a future singularity the transition into a turbulent kind of motion seems  physically inevitable, as the local Reynolds number becomes then very high. That brings  the {\it shear} viscosity concept  back in the consideration, now not in a macroscopic but in a microscopic (local) sense. We expect that there is  established a distribution of eddies over the wave number spectrum.  Most likely this distribution can be taken to be approximately isotropic, implying the existence of an inertial subrange  in which the energy density is $E(k)= \alpha_K \epsilon^{2/3}k^{-5/3}$, where $\alpha_K$ is the Kolmogorov constant, $\epsilon$ the mean energy dissipation, and $k$ the wave number. Ultimately, when the magnitude of $k$ reaches the inverse Kolmogorov length $1/\eta_K = (\epsilon/\nu^3)^{1/4}$  with $\nu$ the kinematic viscosity, the local Reynolds number becomes of order unity and heat dissipation occurs. What we have done above, is to denote the post-transition period $t>t_*$ conceptually as a turbulent region, where the influence from the bulk viscosity has essentially gone away and  where $\gamma_{\rm turb}$ has become positive, without going into further detail as regards the underlying physical transition process.

\section{Conclusions}

Starting from the modified gravity action integral (\ref{4}),  we solved the (00) component of the governing equation (\ref{5}) inserting the scaling relation in Eq.~(\ref{10}), $H=H_0/X$. The bulk viscosity $\zeta$ was assumed in the form (\ref{15}), generalizing the Einstein-case value (\ref{11}) frequently used in the literature. It turned out that the mentioned ansatz (\ref{15}) permitted solutions in the form (\ref{21}), corresponding to future Big Rip singularities. This is thus a Big Rip scenario induced by the bulk viscosity.

 In Sect. 4 it was discussed how the future singularity can nevertheless  in principle be avoided, if one allows for a future phase transition in the cosmic fluid where the influence from viscosity goes to zero. Modified, or conventional, cosmology behave in this sense  essentially in the same way.

 Finally, one may ask to what extent the above theory can be generalized to more complicated forms of the modified Lagrangian term than the simple power-law form given in Eq.~(\ref{3}). Such a general expression for $F(R)$ would then have to be inserted into the  field equation (\ref{2}), and thereafter to be combined with the energy conservation equation (\ref{8}). The general case seems difficult to handle, but there may be special cases that are mathematically tractable and of physical interest. Such an investigation is however outside the scope of the present paper. At present, the scaling laws like Eq.~(\ref{21}) appears to be closely linked to our basic ansatz (\ref{3}).

\bibliographystyle{mdpi}
\makeatletter
\renewcommand\@biblabel[1]{#1. }
\makeatother

\end{document}